# A Bias Aware News Recommendation System

Anish Anil Patankar, Joy Bose, Harshit Khanna

Samsung R&D Institute, Bangalore, India
anish.p@samsung.com, joy.bose@samsung.com, h.khanna@samsung.com

**Abstract.** In this era of fake news and political polarization, it is desirable to have a system to enable users to access balanced news content. Current solutions focus on top down, server based approaches to decide whether a news article is fake or biased, and display only trusted news to the end users. In this paper, we follow a different approach to help the users make informed choices about which news they want to read, making users aware in real time of the bias in news articles they were browsing and recommending news articles from other sources on the same topic with different levels of bias. We use a recent Pew research report to collect news sources that readers with varying political inclinations prefer to read. We then scrape news articles on a variety of topics from these varied news sources. After this, we perform clustering to find similar topics of the articles, as well as calculate a bias score for each article. For a news article the user is currently reading, we display the bias score and also display other articles on the same topic, out of the previously collected articles, from different news sources. This we present to the user. This approach, we hope, would make it possible for users to access more balanced articles on given news topics. We present the implementation details of the system along with some preliminary results on news articles.

**Keywords.** Fake News; Recommendation System; Bias Detection

## 1   Introduction

Political polarization and fake or biased news have been topics of interest in recent times [1]. A number of solutions have been proposed to detect and flag fake news. Many of these solutions seek to prioritize trusted news sources at the server level, for example efforts by Google to partner with fact checking networks [2]. Another is to poll users for trust in news sources, as in Facebook's approach [3]. One criticism of such top down approaches is that they restrict user choice and is tantamount to censorship [4], since the user should be able to decide what kind of news articles they wish to read. In this paper, we present an alternative solution that seeks to give more choice to the end user to make an informed decision.

Our system first informs the reader how much the news article is biased by calculating and displaying a bias score to benchmark the current article with others, similar to the approach by Patankar et. al [5, 6]. After this, it offers recommendations from other news sources on the same topic as the current news article.

In our solution, we use the results of a Pew Research study [7] of the readers of some major English language news sources, positioning the news sources a spectrum from conservative to liberal on the basis of inclinations of political views of their readers. Our system extracts news articles on the same topic as the currently read article from a few of the same news sources from the Pew study. We then display links of the articles taken from the news sources to the user in the form of a graph with the political inclination of the news source on the X axis and the bias score of the news article on the Y axis. This, we hope, would help the user make an informed decision of what kind of news article they wish to read. Our system gives an opportunity to the reader to explore articles that are different from their common political preferences.

The rest of this paper is structured as follows: in the following section we survey related work in bias detection and making a bias aware news recommendation system. Section 3 contains the outlines of our method, along with some implementation specific details. Section 4 presents the results of a user study to analyze how actual users liked the system. Section 5 concludes the paper.

## 2 Related Work

There are a number of works related to detecting bias in articles. Many of them use detection of sentiment bearing words [9, 10] to identify if an article is biased. Recasens et. al. [11] used a method to detect biased language, using a dictionary of words used by Wikipedia editors to ensure articles conform to Neutral Point of View (NPOV) rules.

A few university students developed a plugin called OpenMind [14, 15] to counter fake news, as part of a recent Yale university hackathon. It gives the users feedback on their political biases in their reading patterns, as well as warn users if the source of the news articles is suspected to be fake. However, as per the details available online, it does not give feedback to the user as to how much biased is the current article they are reading. The methodology they use to detect fake news is also unclear.

A similar function is also claimed by BS Detector [16], a Chrome extension that uses a curated list of unreliable news sources to flag online articles as being fake news or otherwise unreliable. This too does not detect and state how much biased the article currently being read is, it is based on reliability of the source of the article rather than the content. Also, they offer no alternative article to read for the same topic. Our system, on the other hand, is aimed at making the user aware of the bias in the current article as well as recommend other articles on the same topic from different news sources, so the user can make an informed decision on what they want to read.

## 3 System Overview

In this section, we describe the steps of our approach for bias aware news recommendations in detail.

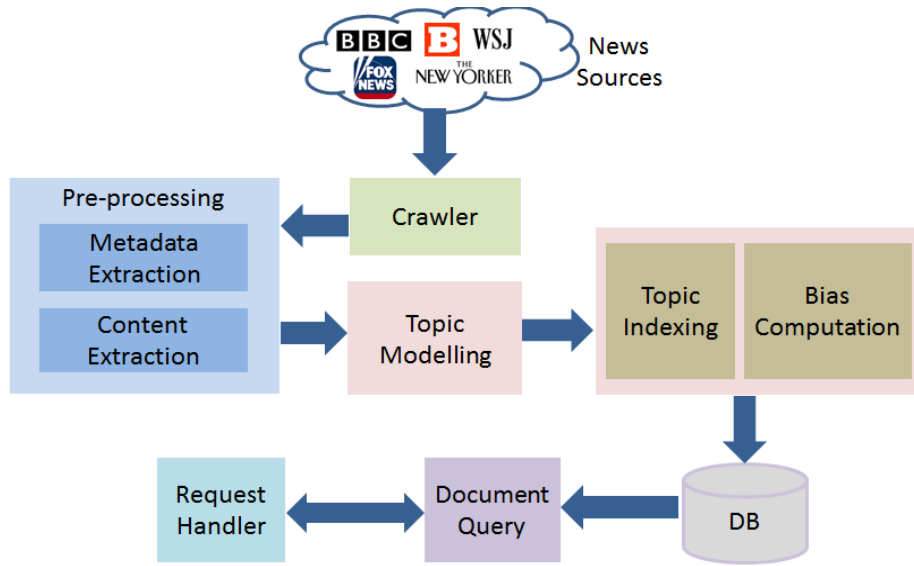

**Fig. 1.** Block diagram of the system for indexing news articles from different sources, along with their bias scores

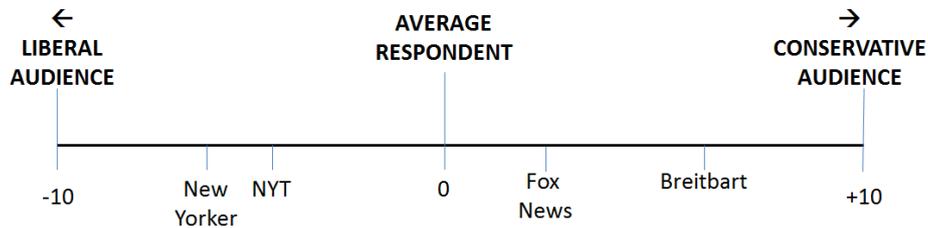

**Fig. 2.** Extract from the Pew Research study on the audience of selected news sources. Adapted from [4]

### 3.1    Crawling news articles

The first step in our system is to crawl news content from a variety of news sources. The block diagram of the system is given in Fig. 1.  We used selected news sources from the Pew Study [7], selecting a mix of sources with conservative and liberal readers, as well as mainstream and non-mainstream media outlets as per Wikipedia [8], where the top 6 media outlets are defined as mainstream. A graph of the scores of selected news articles from the Pew study is shown in Fig. 2.

For each of the news sources selected, we crawl their websites for latest content, by fetching the robots.txt file from the new site domain URL, and then parsing the robots.txt file to extract the sitemap or sitemap index. After this, we parse the XML file of the sitemap and extract the URLs with timestamps. We then go through each of the

URLs to extract the title and text content of the body of the webpage, as well as tags from the title if available.

Based on the availability of the robots.txt and sitemap files, we selected the following five media sources: New Yorker (liberal and non-mainstream), New York Times (liberal and mainstream), BBC (liberal and mainstream), Fox News (conservative and mainstream), and Breitbart (conservative and non-mainstream).

After collecting the text of the articles, we calculated and added the bias score for each article, as mentioned in the next subsection. We use this index to query similar articles from the crawled data.

For indexing, we used the semantic indexing system used by Sailesh et al [8]. The system uses an LDA based [13] indexing and matching system. For each article, the system computes the topic distribution within the article. After that, it computes the cosine distance between two articles on the basis of the topic distributions, and on that basis builds a semantic association index.

### 3.2 Computing a bias score for the news articles

In this subsection, we describe the system for computing a bias score for an online news article. As mentioned, we use the system similar to the one used by Patankar et. al [5, 6]. The system uses the non-NPOV lexicon from Wikipedia as created by Recasens [11]. Fig. 3 shows the architecture of the bias detection system for a given news article. The article bias score is found by totaling the biased sentences (containing one or more words from the bias lexicon) and dividing by the total number of sentences in the news article.

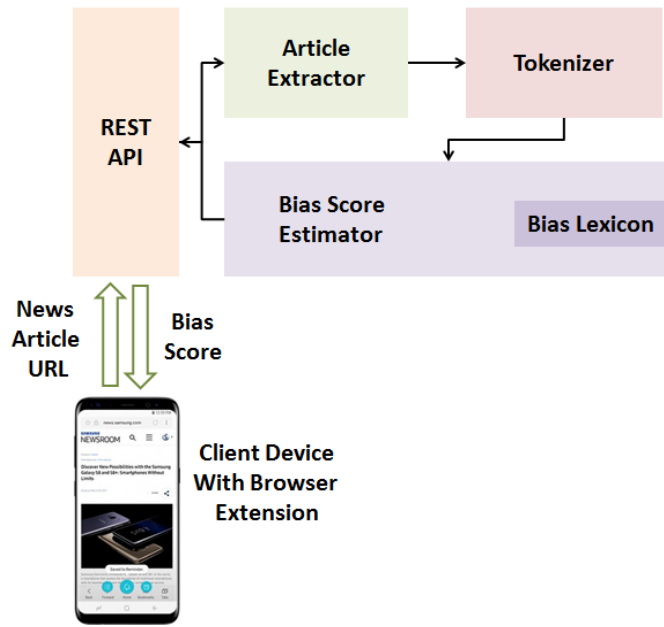

**Fig. 3.** High level architecture of the bias score computation system for one article.

### 3.3 RESTful service for bias aware news recommendations on the same topic

Our system has a web browser extension that acts as a client to the web service. It send the article URL to the service which returns the resulting bias score for the article as well as the URL recommendations. On the server, the URL is fetched, the URL content is extracted and queried in our database for matching articles from different news sources. On the client, the browser extension queries the RESTful service and show the results to the user and allows the user to choose the article.

### 3.4 Implementation for a desktop browser

Figures 4-6 give screenshots of the implementation for a desktop browser. On clicking the browser extension icon to the right of the address bar, the system displays a graph showing the current bias score on the Y axis and the news sources on the X axis. The current article bias score is shown in red, while the icons of the news sources represent links to news articles from those sources on the same topic. On hovering on the links, the user can see the news article topic along with the bias score.

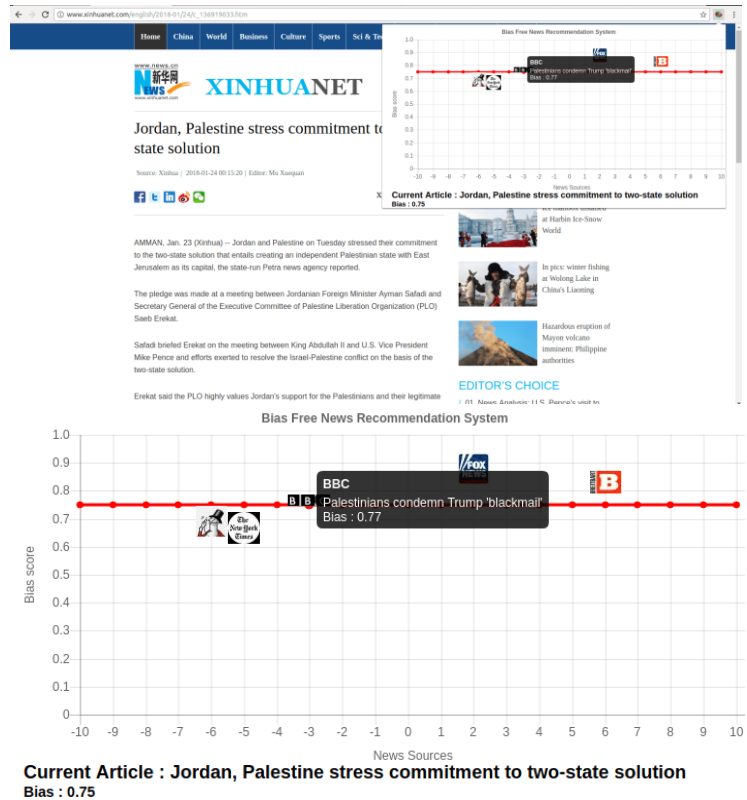

**Fig. 4.** Screenshots of the popup showing the related news articles from different sources on the same topic, on a desktop web browser for a news article.

(a)

(b)

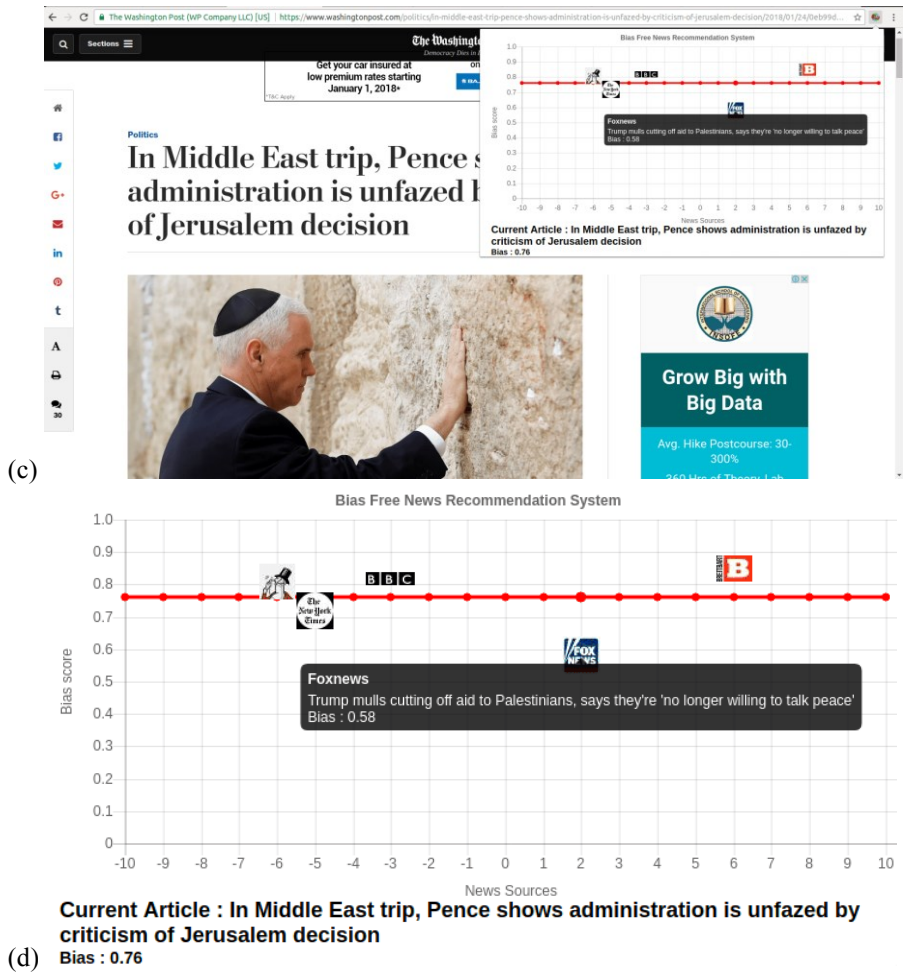

**Fig. 6.** Screenshots showing the recommended news articles from different sources on the same topic, plotted along with the bias scores, on a desktop web browser for a few news articles.

### 3.5 Implementation on a mobile phone

We also implemented the system on a web browser in the Android mobile phone, using the semantic browsing framework described in [8]. Here, instead of displaying the links along with the bias scores in a table, we displayed the icons of the news sources on the same topic at the bottom of the web browser. The user flicks the news article left or right with the finger to see and read other articles from different news sources on the same topic. Fig. 7 shows the screenshots of related articles on the same topic for a mobile web browser.

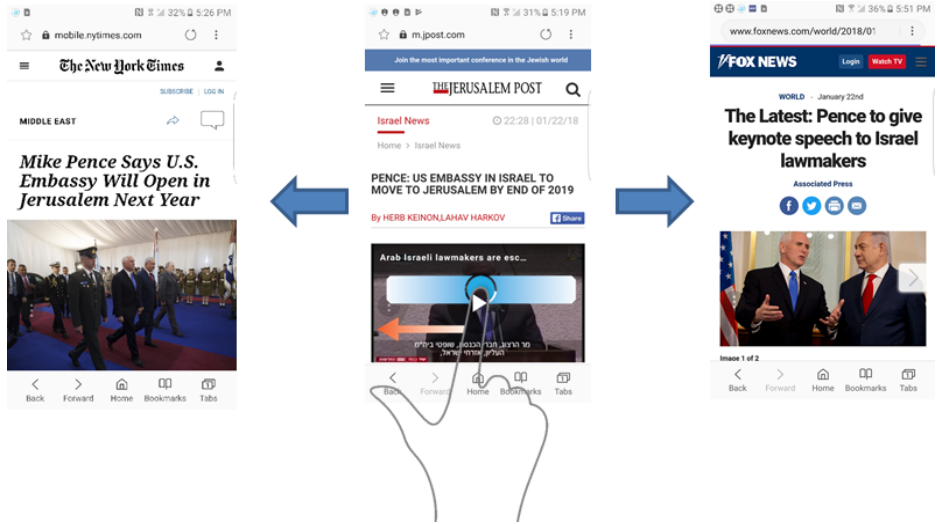

**Fig. 7.** Screenshots of the solution on a mobile phone, where the user has to swipe left or right with their finger to see articles from other news sources on the same topic, utilizing the semantics based browsing framework from [8].

## 4 Experimental setup and results

We implemented the system on a Chrome web browser on the desktop with a browser extension, as well as on the default browser on a Samsung Galaxy S6 mobile phone.

### 4.1 Correlation of algorithm and user generated bias scores

We studied the bias scores obtained for the current article and recommended articles, for a variety of news articles on relevant current topics.

We conducted a small user study with 5 users to gauge (a) the accuracy of the system and (b) its usefulness and ease of use. We gave the users 3 sets of 5 articles each (each set of 5 articles being from the same topic but from different news sources), and asked them to rank the articles for the level of perceived bias. Table 2 gives the bias scores generated by our algorithm along with the URL recommendations (with bias scores) for the three sets of articles.

Then we compared the user generated rankings of bias with those generated by our algorithm. We found that the rankings of bias correlated well with the bias scores generated by our algorithm.

### 4.2 Subjective reviews by reviewers

We also collected the subjective reviews by the users, on the overall usability of the system. Most of the users thought the system could be useful, though a couple of users thought they were not politically aware enough to judge how much biased the articles were. Since the number of users is too small to be meaningful, in future we will repeat the study with more users.

**Table 2.** Bias scores along with URL recommendations for a few sample articles

| Query URL and bias | Matching article URL | Bias Score |
|---|---|---|
| https://www.wsws.org/en/articles/2018/01/06/paki-j06.html<br><br>Bias : 0.82 | https://www.newyorker.com/news/news-desk/the-tapi-pipeline-and-paths-to-peace-in-afghanistan | 0.72 |
| | https://www.nytimes.com/2018/01/09/opinion/pakistan-trump-aid-en-gage.html?mtrref=undefined&gwh=7973905D6242465D0C400CC5DA93A330&gwt=pay&assetType=opinion | 0.68 |
| | http://www.bbc.com/news/world-us-canada-42574139 | 0.92 |
| | http://www.foxnews.com/story/2007/09/21/heritage-foundation-leveling-with-pakistan-on-afghanistan.html | 0.81 |
| | http://www.breitbart.com/national-security/2017/03/28/pakistan-islamabad-constructing-fence-along-afghan-border/ | 0.68 |
| http://www.xinhuanet.com/english/2018-01/24/c_136919033.htm<br><br>Bias : 0.75 | https://www.newyorker.com/news/news-desk/a-significant-deal-between-the-u-s-and-israel | 0.69 |
| | https://www.nytimes.com/2018/01/03/us/politics/trump-tweets-nuclear-but-ton.html?mtrref=undefined&gwh=DD4472067A407A5828213D469AE408A7&gwt=pay | 0.67 |
| | http://www.bbc.com/news/world-us-canada-42553507 | 0.77 |
| | http://www.foxnews.com/opinion/2011/09/16/time-is-now-for-three-state-solution-to-end-israeli-palestinian-conflict.html | 0.88 |
| | http://www.breitbart.com/national-security/2017/03/30/leaders-of-egypt-jordan-and-palestinian-authority-to-meet-with-trump-in-april/ | 0.83 |
| https://www.washingtonpost.com/politics/in-middle-east-trip-pence-shows-administration-is-unfazed-by-criticism-of-jerusalem-decisi-si- | https://www.newyorker.com/news/news-desk/trump-sabotages-his-own-mideast-peace-process | 0.81 |
| | https://www.nytimes.com/2018/01/03/world/middleeast/trump-israel-palestinians-twit-ter.html?mtrref=undefined&gwh=14D4AB4FB34FB4F9B44ABBC9389845DB&gwt=pay | 0.72 |
| | http://www.bbc.com/news/world-us-canada-42402350 | 0.82 |
| | http://www.foxnews.com/politics/2018/01/02/trump-mulls-cutting-off-aid-to-palestinians-says-theyre-no-longer-willing- | 0.58 |
| on/2018/01/24/0eb99d30-0125-11e8-9d31-d72cf78dbeee_story.html?utm_term=.9b57a537b857<br><br>Bias : 0.81 | to-talk-peace.html | |
| | http://www.breitbart.com/jerusalem/2017/05/01/abbas-preps-trump-meeting-week/ | 0.85 |

## 5   Conclusion and future work

In this paper, we have described a system to recommend related URLs from the same topic from news sources with a variety of political biases. We hope that such a system can help to enable users to be more aware of bias in news articles they are currently reading. Our system offers a compromise between keeping user autonomy and the need to warn the users of biased or fake news.

In future, we aim to conduct a thorough user study to gauge the effectiveness of the system. We also aim to develop a more intelligent system for bias estimation using machine learning techniques to improve our simple dictionary based approach, as well as explore sentiment analysis along with bias to develop a composite score for the article.